\documentclass{elsart}

\usepackage[square]{natbib}

\usepackage{graphicx}
\graphicspath{{figures/}}

\usepackage{amssymb}
\journal{Nucl. Instr. and Meth. in Phys. Res. A}

\def\PANDA{$\overline{\mbox{P}}${ANDA}\ }
\def\d{\mathrm{d}}

\begin{document}

\begin{frontmatter}

\title{Measurement of Propagation Time Dispersion in a Scintillator}

\author{P.~Achenbach\corauthref{cor}},
\ead{patrick@kph.uni-mainz.de}
\author{C.~{Ayerbe Gayoso}},
\author{J.~Bernauer},
\author{R.~B\"ohm},
\author{M.O.~Distler},
\author{L.~Doria},
\author{J.~Friedrich},
\author{H.~Merkel},
\author{U.~M\"uller},
\author{L.~Nungesser},
\author{J.~Pochodzalla},
\author{S.~{S\'anchez Majos}\thanksref{PhD}},
\author{S.~Schlimme},
\author{Th.~Walcher}, \and
\author{M.~Weinriefer}

\address{Institut f\"ur Kernphysik, Johannes Gutenberg-Universit\"at
    Mainz, Germany}

\thanks[PhD]{Part of doctoral thesis.}
\corauth[cor]{Corresponding author. Tel.: +49--6131--3925831;
fax: +49--6131--3922964.}

\begin{abstract}
  One contribution to the time resolution of a scintillation detector
  is the signal time spread due to path length variations of the
  detected photons from a point source. In an experimental study a
  rectangular scintillator was excited by means of a fast pulsed
  ultraviolet laser at different positions along its longitudinal
  axis. Timing measurements with a photomultiplier tube in a detection
  plane displaced from the scintillator end face showed a correlation
  between signal time and tube position indicating only a small
  distortion of photon angles during transmission. The data is in good
  agreement with a Monte Carlo simulation used to compute the average
  photon angle with respect to the detection plane and the average
  propagation time. Limitations on timing performance that arise from
  propagation time dispersion are expected for long and thin
  scintillators used in future particle identification systems.
\end{abstract}

\begin{keyword}
   timing detectors \sep
   time-of-flight spectroscopy \sep
   particle identification
\end{keyword}

\end{frontmatter}

\section{Introduction}
There has been recent interest in the timing performance of long and
thin plastic scintillation detectors with demands on the time
resolution of 100\,ps and below, e.g.\ for the future \PANDA
experiment at {\sf FAIR}~\cite{PANDA}. An existing experiment with a
challanging scintillator system is COMPASS with the installation of
the 2.8\,m long and 4\,mm thin recoil detector under
way~\cite{COMPASS}. This paper shows that the detector performance
required by these experiments is close to the fundamental limit that
arises from the statistical fluctuations in the process of the signal
generation.

At \PANDA, the identification of charged particles with momenta up to
several GeV$/c$ will be performed by the detection of internally
reflected Cherenkov (DIRC) light~\cite{DIRC-GSISciRep05}. Any
measurement of the time-of-propagation (TOP) in the DIRC counter needs
a precise time reference and can be combined with the time-of-flight
(TOF) information from long scintillator slabs in front of the
radiator barrel of the DIRC\@. Within the almost fully hermetic \PANDA
detector such a scintillator array needs to be as thin as physically
possible. At COMPASS, it is the need for a low momentum proton
detection that requires the scintillators to be very thin.

The timing properties of scintillators are usually defined in terms of
a coincidence time resolution. In a typical laboratory measurement the
time difference between the discriminated signals of two
photomultipliers (PMTs) placed at the extremes of a sample of
scintillator is measured for minimum ionising particles crossing the
scintillator at its centre. For small counters of 45\,cm length and
20\,mm thickness a time resolution reduced for one PMT of the order of
FWHM $\approx$ 150\,ps has been achieved by the authors in the
laboratory. For long scintillators there are several effects degrading
the time resolution, namely the light attenuation and the consequences
of the scintillator acting as a light guide with corresponding
propagation time dispersion.

First photoelectron timing errors have been evaluated for slow
scintillators (BGO, $\tau_\mathrm{decay} \approx 300$\,ns) and larger
amplitudes (number of photoelectrons $N> 30$)
previously~\cite{Petrick1991,Clinthorne1990}, but not for fast plastic
scintillators and small amplitudes, where the achievable timing
performance depends also on path length variations. The purpose of
this paper is to discuss expressions for the timing error which
include propagation time dispersion and to verify experimentally the
the correlation between signal time and average photon angle at the
read-out end of a scintillator.

\section{The Origin of Time Jitter in Scintillator Timing}
One fundamental limit on the time resolution of scintillation counters
comes as a consequence of the statistical processes involved in the
generation of the signal. Post and Schiff~\cite{PostSchiff1950} have
first discussed such limitations. In general, the voltage pulse of a
photomultiplier, $ V_\mathrm{PMT}$, can be expressed as a linear
superposition of $N$ single photoelectron pulses, $v_i$, whose
amplitude is allowed to vary due to gain fluctuations, $g_i$, governed
by a measurable probability distribution. The pulses arrive at
individual times due to the time spread in the energy transfer to the
optical scintillator levels, $t_\mathrm{dep}$, the decay time of the
light emitting states, $t_\mathrm{emit}$, the propagation time,
$t_\mathrm{prop}$, and the transit time, $t_\mathrm{TT}$, being the
time difference between photo-emission at the cathode and the arrival
of the subsequent electric signal at the anode. In addition, there is
white Gaussian electronic noise, $w(t)$. We incorporate these
processes into a general model: $V_\mathrm{PMT}(t)=
\sum_{i=1}^{N}{g_iv_i(t - (t_\mathrm{dep} +
  t_\mathrm{emit}+t_\mathrm{prop}+t_\mathrm{TT})_i)} + w(t)$, where
$N$ fluctuates from one pulse to another. In a semi-classical model
the probability for observing $N$ photoelectrons during a time
interval $T$ is given by a Poissonian distribution $P(\overline{N},
N)$ with $\overline{N}$ being the mean number of photoelectrons per
pulse.

The following calculation is focused on the limitations arising from
the contribution of the path length variation. Central to the
calculation is the relation between initial axial angle, $\theta$, and
propagation time: $t= Ln/(c\cos\theta)$, for a point like source of
light placed at a distance $L$ from the end face of a cylindrical
scintillator with refractive index $n$ in a medium of refractive index
$n_\mathrm{ext}$, where $\theta_\mathrm{max}$ represents the
complement of the total internal reflection angle,
$\theta_\mathrm{max}= \arccos{n_\mathrm{ext}/n}$. In the case of a
rectangular shaped scintillator the relation is still valid as can be
seen from the following consideration: in any reflection the change in
the velocity vector of a particular photon takes place in the
direction perpendicular to the face it hits. The component of this
vector along the longitudinal axis remains unchanged during the whole
motion and therefore the propagation time only depends on the initial
axial angle, distance to the PMT, and index of refraction.
Consequently, a detector displaced by a distance $d$ from the end face
of the scintillator would link in one linear dimension the angle and
propagation time with an accuracy given by $\sigma_\theta\approx
\sqrt{t_x^2/12 + a_x}/d$, where $t_x$ is the thickness of the
scintillator and $a_x$ the aperture of the detector.

For a quantitative description including extended light sources, light
attenuation and refraction a full photon tracking simulation with the
correct geometry of the set-up and a good model of scintillation light
excitation and emission was needed.  A general purpose Monte Carlo
program~\cite{Gentit2002} simulating light propagation was used to
compute the average photon angle with respect to the detection plane
and the average propagation time.

Experimentally, the quality of a scintillator surface limits the
conservation of photon angles. Small geometrical inhomogeneities can
have a strong impact on the transmission of photons and then the
correlation between detection position and propagation time is heavily
affected. There is no straight forward method for the estimation of
the parameters necessary for an appropriate simulation of the surface
quality. In fact, a Monte Carlo refractive index matching technique
was developed only very recently to determine these
parameters~\cite{Wahl2007}. It was therefore mandatory to perform an
experiment with a sample of plastic scintillator in order to
investigate empirically the mentioned correlation.

In an experiment the discriminators and any noise in the electronic
circuits contribute to the time spread. Further, the response time of
discriminators may shift with signal amplitude ("time walk"). Time
walk effects are relevant if a large range of amplitudes is
discriminated and can get corrected by various means, either in
hardware or software.

\begin{figure}
  \centerline{\includegraphics[width=0.48\textwidth]{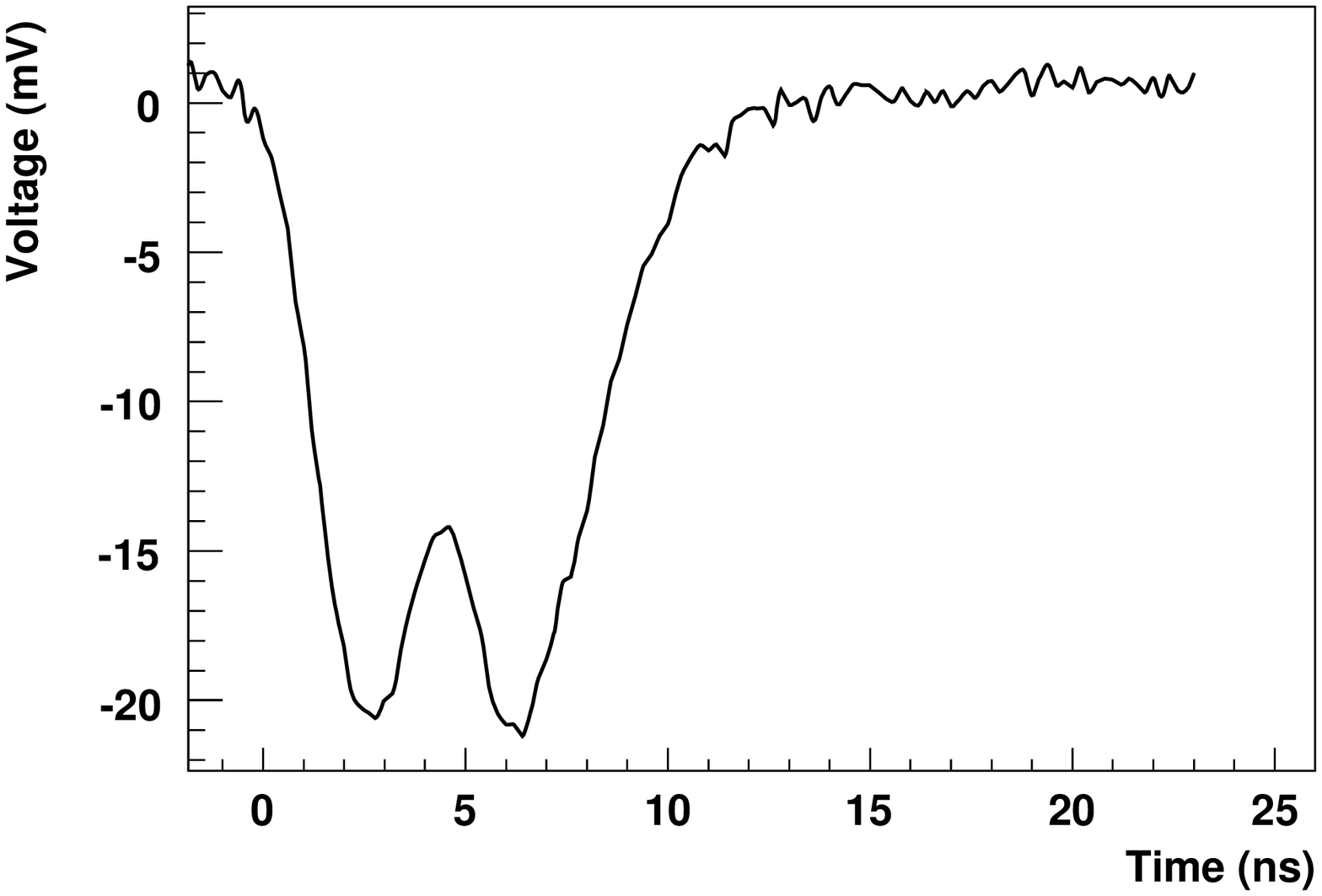}
              \includegraphics[width=0.48\textwidth]{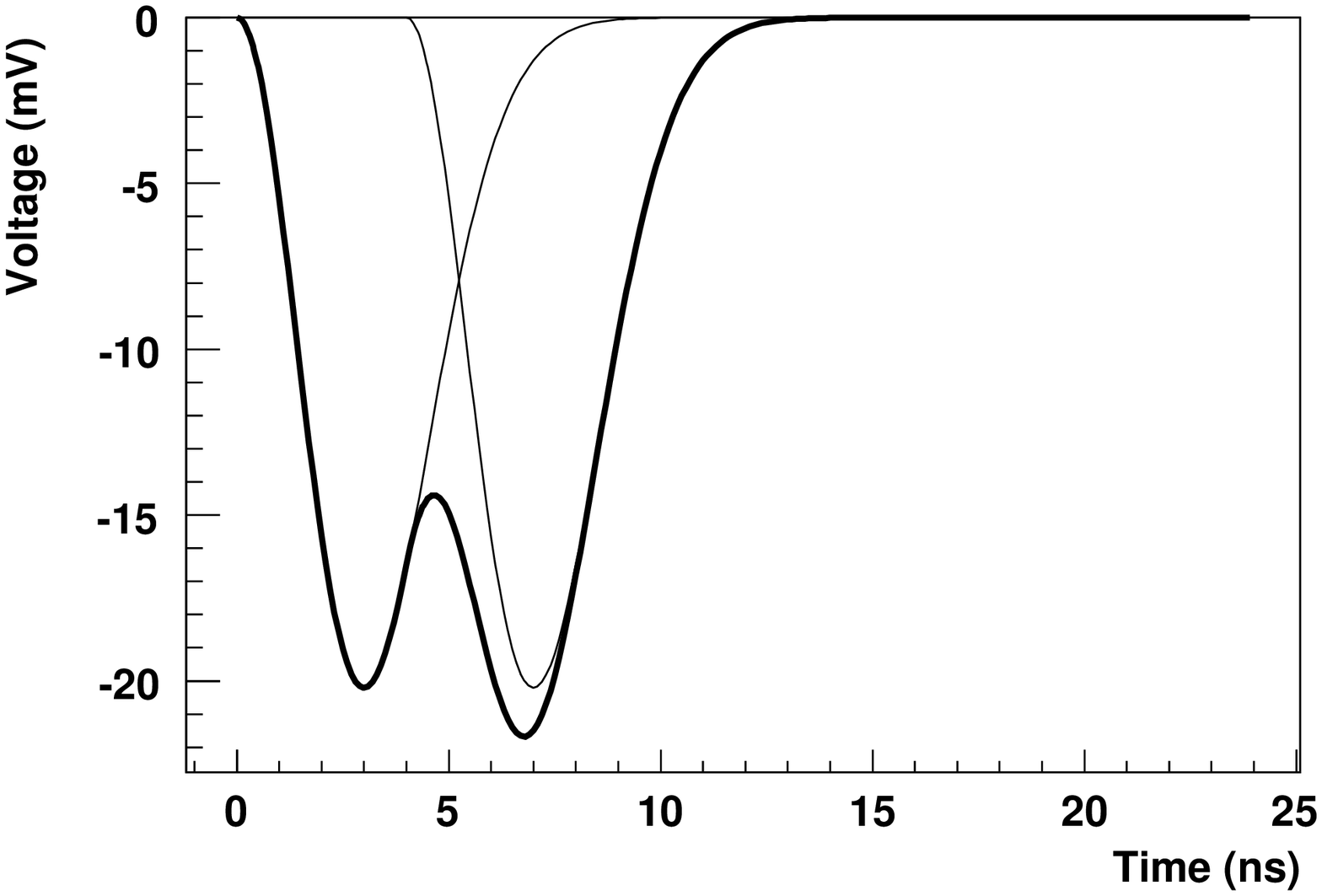}}
  \caption{Left: Typical PMT voltage pulse for low light intensities
    as measured with a digital oscilloscope.  Right: Typical voltage
    pulse calculated by a Monte Carlo method for two single 
    photoelectron pulses separated by 5\,ns.}
  \label{fig:signal}
\end{figure}

It is a standard in nuclear physics to use constant fraction
discriminators (CFD) and leading edge discriminators (LED), the latter
showing a characteristic time walk. In many cases constant fraction
timing provides the best time resolution with scintillation counters.
The situation is significantly different when the number of photons is
small. In this case the sampling of the arrival time distribution is
poor and the signal shape is very variable. Fig.~\ref{fig:signal}
shows the trace of a PMT voltage pulse for low light intensities as
measured with a digital oscilloscope and the result of a Monte Carlo
calculation for $N=$ 2 photoelectrons. The voltage pulse of a single
photoelectron was modelled by $v_\mathrm{spe}(t)= gc t^2
\e^{-{t^2}/{\tau^2}}\! / \int_{0}^{\infty} t^2 \e^{-{t^2}/{\tau^2}}\!
\d t$, where $g$ is the gain of the PMT, $c$ is the conversion factor
for charge to voltage and $\tau$ has been chosen to be 4\,ns in order
to match the signal of the PMT used in the experimental set-up. A
large distribution of pulse shapes reflecting the different arrival
times of the individual photons is possible and the confounding effect
of the overlapping single photon responses leads to problems with
leading edge and peak detectors. Thus, the timing performance in the
pile-up case degrades considerable from single photon timing.

\section{Calculation of Fundamental Limits from Propagation
  Time Dispersion}
The spread of propagation times is given by $\Delta t_\mathrm{prop}= L
n/c(\cos^{-1}{\theta_\mathrm{max}}-1)$ for the minimum and maximum
propagation times of meridional rays, $t_\mathrm{min}= Ln/c$ and
$t_\mathrm{max}= Ln/(c\cos\theta_\mathrm{max})$. The probability
density function of the propagation times of photons, $\d N/(N \d t)$,
produced by an event at $t= 0$, can be calculated from the angular
distribution of photons inside the scintillator, $\d N= 2\pi\,
\d\cos\theta$. It follows that $\d N/\d t= -2\pi\, L n/(ct^2)$ and $N=
\int_{t_\mathrm{min}}^{t_\mathrm{max}} (\d N/\d t) \d t'= 2\pi\,
(\cos\theta-1)$. Finally, the probability density for the arrival
times at the photocathode due to path length variations is
$P_\mathrm{prop}(t)= \d N/(N \d t)= Ln/((\cos\theta-1)ct^2)$.

The arrival time probability density function for a realistic
scintillator can be folded in as follows: $P(t)= \int_{0}^{\infty}
I(t') P_\mathrm{prop}(t-t') P_\Lambda(t-t') \d t'$, where
$P_\Lambda(t)= \exp{-tc/(n \Lambda)}$ is survival probability
depending on the bulk absorption coefficient, $\Lambda$, in the
material. The distribution of decay times of light emitting states is
included with a light pulse shape of the form: $I(t)= [(e^{-t/\tau_2}
- e^{-t/\tau_1}) / (\tau_2 - \tau_1) + R e^{-t/\tau_3} / \tau_3 ] / (1
+ R)$, where $\tau_1$ and $\tau_2$ are two fast decay constants,
$\tau_3$ corresponds to the decay time of the slow component, and $R$
is the ratio of the slow to fast components. For the parameters of
emission time distribution values of $\tau_1=$ 0.9\,ns, $\tau_2=$
2.1\,ns, $\tau_3=$ 14.2\,ns, and $R=$ 0.27 were chosen. Comparable
values are given in Ref.~\cite{Nam2002}. The probability density
function of the transit time spread of electrons in the
photomultiplier, $P_\mathrm{TTS}$ is not relevant for the following
discussion, since micro-channel plate PMTs provide an alternative with
excellent timing capabilities. Analytical descriptions of the
probability density functions describing the time jitter of
photoelectrons generated at the photocathode have been discussed
before, e.g.\ in~\cite{Moszynski1979}.

\begin{figure}
  \centerline{\includegraphics[width=0.8\textwidth]{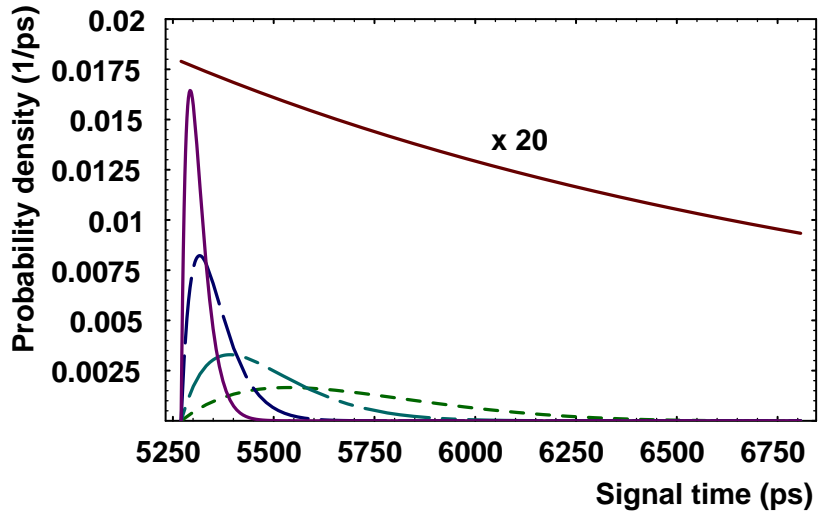}}
  \caption{Calculated distributions of signal times for $L=$ 1\,m,
    $n=$ 1.58, for an increasing number of photoelectrons, $N$, at a
    fixed threshold of $N_\mathrm{thr}=$ 2: $N=$ 5 (solid line), $N=$
    10 (dashed line), $N=$ 25 (dot-dashed line), and $N=$ 50 (dotted
    line). The continuously falling (solid) line is a calculation for
    single photoelectron pulses ($\times$ 20).}
  \label{fig:photonstatistics}
\end{figure}

The probability for a photoelectron to be detected in the interval
between $t_\mathrm{min}$ and $t$ is $p(t)= \int_{t_\mathrm{min}}^t\!
\d N/(N \d t) \d t'$. Assuming a Poissonian distribution for the
probability that a definite number of photoelectrons,
$N_\mathrm{thr}$, has been accumulated leads to the signal time
distribution $S(t)= N \frac{\d N}{\d t} {N-1\choose N_\mathrm{thr}-1}
[p(t)]^{N_\mathrm{thr}-1} [1-p(t)]^{N-N_\mathrm{thr}}$, which is to be
normalised for getting the probability density function.  Calculated
distributions for signal times with $L=$ 1\,m, $n=$ 1.58, for an
increasing number of photoelectrons in the pulse, $N=$ \{5, 10, 25,
50\} and $N_\mathrm{thr}=$ 2, as well as for single photoelectron
pulses ($\times$ 20) are shown in Fig.~\ref{fig:photonstatistics}.

\begin{figure}
  \centerline{\includegraphics[width=0.8\textwidth]{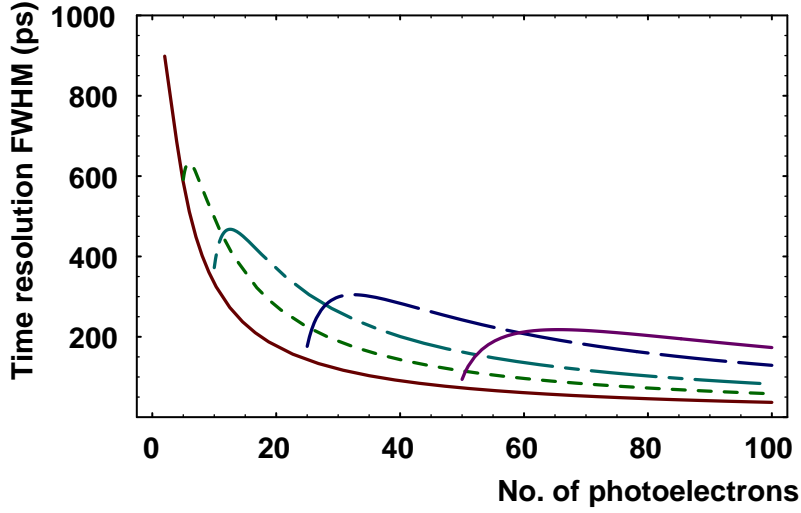}}
  \caption{Calculated limits on time resolution (FWHM) for $L=$ 1\,m,
    $n=$ 1.58, as a function of the number of photoelectrons, $N$, at
    thresholds of $N_\mathrm{thr}=$ 2 (solid line following a
    FWHM~$\propto 1/N$ dependence), $N_\mathrm{thr}=$ 5 (dotted line),
    $N_\mathrm{thr}=$ 10 (dot-dashed line), $N_\mathrm{thr}=$ 25
    (dashed line), and $N_\mathrm{thr}=$ 50 (solid line).}
  \label{fig:resolutionlimits}
\end{figure}

The limit on time resolution arising from statistics of photon
detection can now be calculated using the relations for the
expectation values $\langle t_p \rangle=
\int_{t_\mathrm{min}}^{t_\mathrm{max}}\!t S \d t$ and $\langle t_p^2
\rangle= \int_{t_\mathrm{min}}^{t_\mathrm{max}}\! t^2 S \d t$, and the
variance of the signal time $\mathrm{Var}(t_p)= \langle t_p^2 \rangle
- \langle t_p \rangle^2$. The time resolution (FWHM) for $L$= 1\,m,
$n$= 1.58 as a function of $N$ is shown in
Fig.~\ref{fig:resolutionlimits} for a range of thresholds,
$N_\mathrm{thr}$= \{2,5,10,25,50\}. The curve for $N_\mathrm{thr}$= 2
follows a FWHM~$\propto 1/N$ dependence.

\section{Experimental Set-up}
\begin{figure}
  \includegraphics[width=\textwidth]{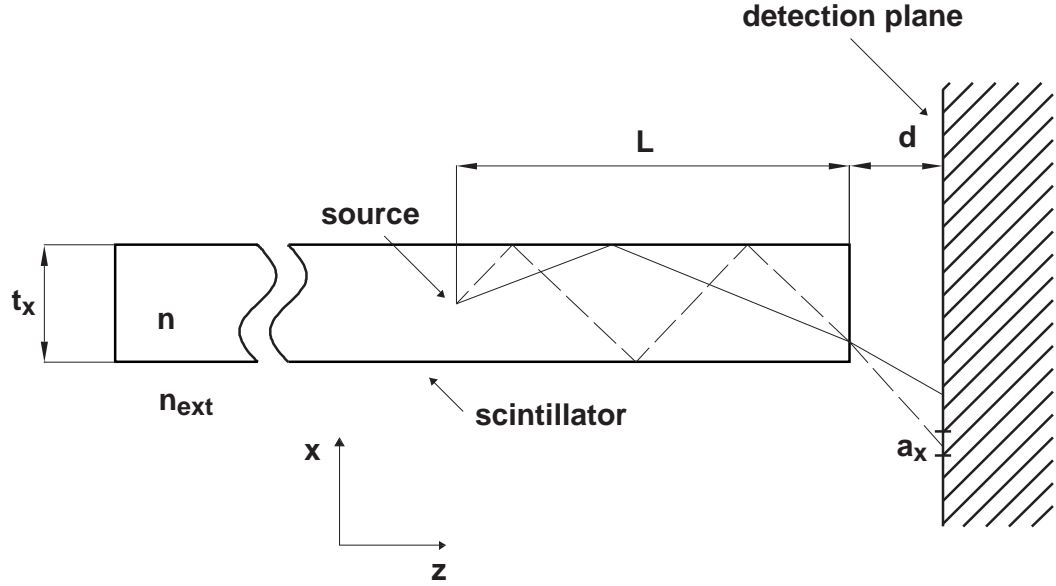}
  \caption{Schematic representation of the arrangement of scintillator
    and detection plane at a distance $d$. The paths of two photons
    leaving the scintillator at the same point are shown. Typical
    parameters for the experiments were $a_{x}= a_{y}=$ 3\,mm, $L=
    100$\,cm, $d= 10\,$mm, $t_{x}= 32$\,mm, and $t_{y}= 10$\,mm.}
  \label{fig:setup}
\end{figure}

\begin{figure}
  \centerline{\includegraphics[width=0.6\textwidth]{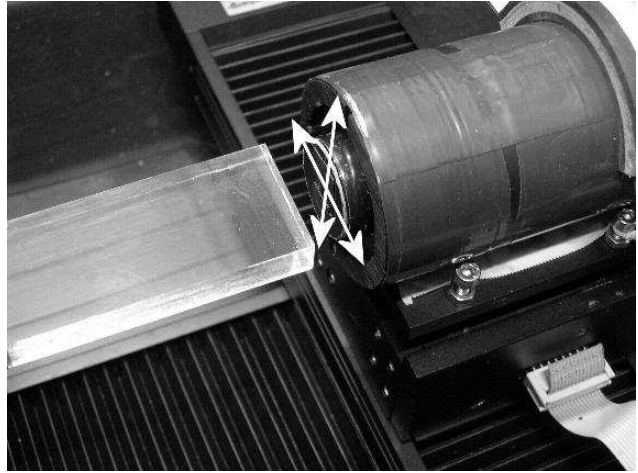}}
  \caption{The photograph shows the crossed positioning units and the
    PMT equipped with a plastic mask with a square aperture of 3\,mm
    $\times$ 3\,mm. The PMT motion takes place in the plane
    perpendicular to the longitudinal axis of the scintillator.}
  \label{fig:photo}
\end{figure}

The basic idea was to measure the average photon angle with respect to
the detection plane at a distance $d=$ 1\,cm.  A schematic
representation of the arrangement of scintillator and detection plane
is given in Fig.~\ref{fig:setup}. The trajectories of two photons
leaving the scintillator in the same point are shown. A BC-408
scintillator from {\sf Bicron} of dimensions $32 \times 10 \times
2000$\,mm$^3$ was used. The scintillator is characterised by a decay
time of $\tau_\mathrm{decay}=$ 2.1\,ns, a small admixture of a slower
decay time, and a refraction index of $n=$ 1.58. It was important for
the conservation of photon angles that the faces of the scintillator
were not wrapped or covered. A minimum contact area of the bar at only
two edges was accomplished by placing the scintillator along the
longitudinal axis of a black metallic tube of the proper dimensions.
The two tubes of 1\,inch diameter were of type R4998 from {\sf
  Hamamatsu} with a fast time response and a linear focused dynode
structure of 10 stages, leading to a transit time spread of only
$\sigma_\mathrm{TTS}=$ 160\,ps. At one end the reference PMT was fixed
keeping contact with the scintillator by a suitable set of springs. At
the other end a set of two linear positioning stages permitted the
scanning of the detection plane behind the scintillator, see
Fig.~\ref{fig:photo}. A thin plate with a 3\,mm wide aperture was
mounted in front of the PMT window to limit the acceptance. A
$^{90}$Sr source and a fast pulsed ultraviolet laser from {\sf Horiba
  Jobin Yvon} were used for the measurements. The laser had a pulse
duration FWHM $< 200$\,ps. The signals were digitised by a charge
integrating analogue-to-digital converter ({\sf LeCroy} 1885F,
50\,fC$/$count) and by a {\sf LeCroy} time-to-digital converter ({\sf
  LeCroy} 1875, 25\,ps$/$count). The TDC digitisation corresponds to
an accuracy of 25$/\sqrt{12}$\,ps $\approx$7\,ps.

A full scanning of the amplitudes was performed by means of filters
placed in front of the movable PMT. The constant fraction
discriminators showed little time walk ($\Delta t/\Delta Q <
0.7$\,ps$/$pC). It was fitted and used for the correction of the TDC
information.

\section{Discussion of Results}
\begin{figure}
  \centerline{\includegraphics[width=0.8\textwidth]{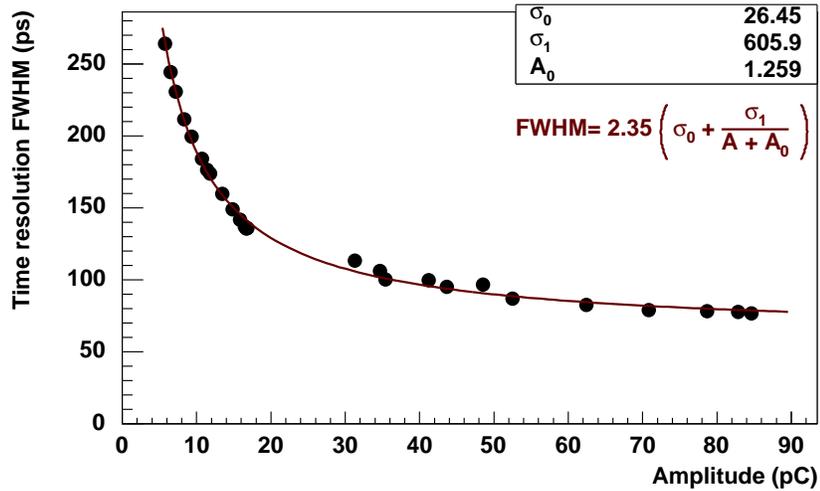}}
  \caption{Measured time resolution (FWHM) obtained with a BC-408
    scintillator and an UV laser of varying primary light intensity as
    a function of the amplitude, $A$. The function FWHM $\propto
    \sigma_0+\sigma_1/(A+A_0)$ has been fitted to the data.}
  \label{fig:timeResVSint}
\end{figure}

A time resolution of FWHM $\approx 130-260$\,ps was achieved for low
amplitudes, see Fig.~\ref{fig:timeResVSint}. With increasing amplitude
the resolution improved until it levelled off at about FWHM $\approx
80$\,ps, demonstrating that at high intensities the photon statistics
is no longer decisive and the resolution is supposedly dominated by
electronic noise. A function of the form FWHM $\propto
\sigma_0+\sigma_1/(A+A_0)$ has been fitted to the data under the
assumption that the detector response becomes Gaussian as the number
of photoelectrons increases.

\begin{figure}
  \includegraphics[width=0.48\textwidth]{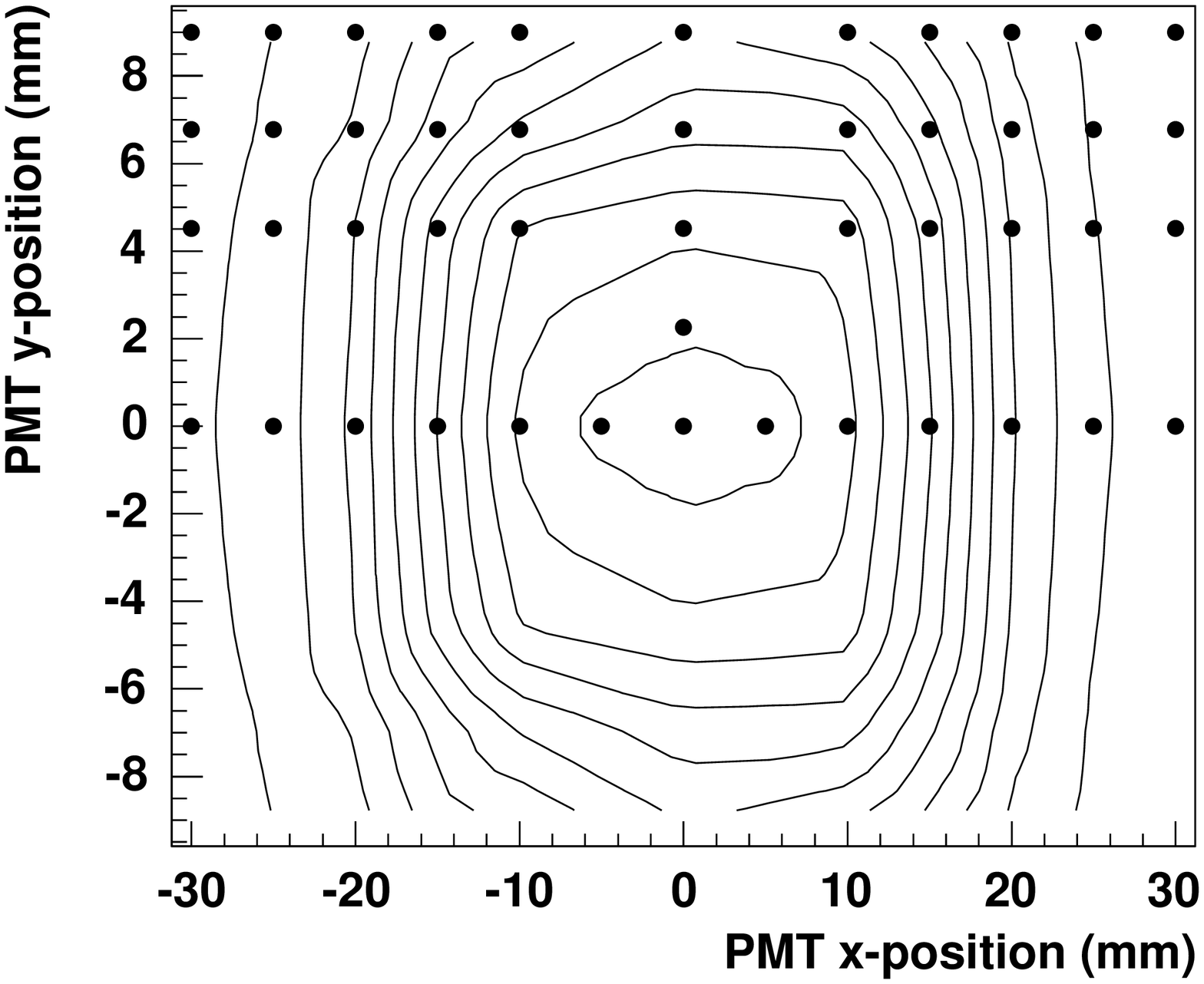}\hfill
  \includegraphics[width=0.48\textwidth]{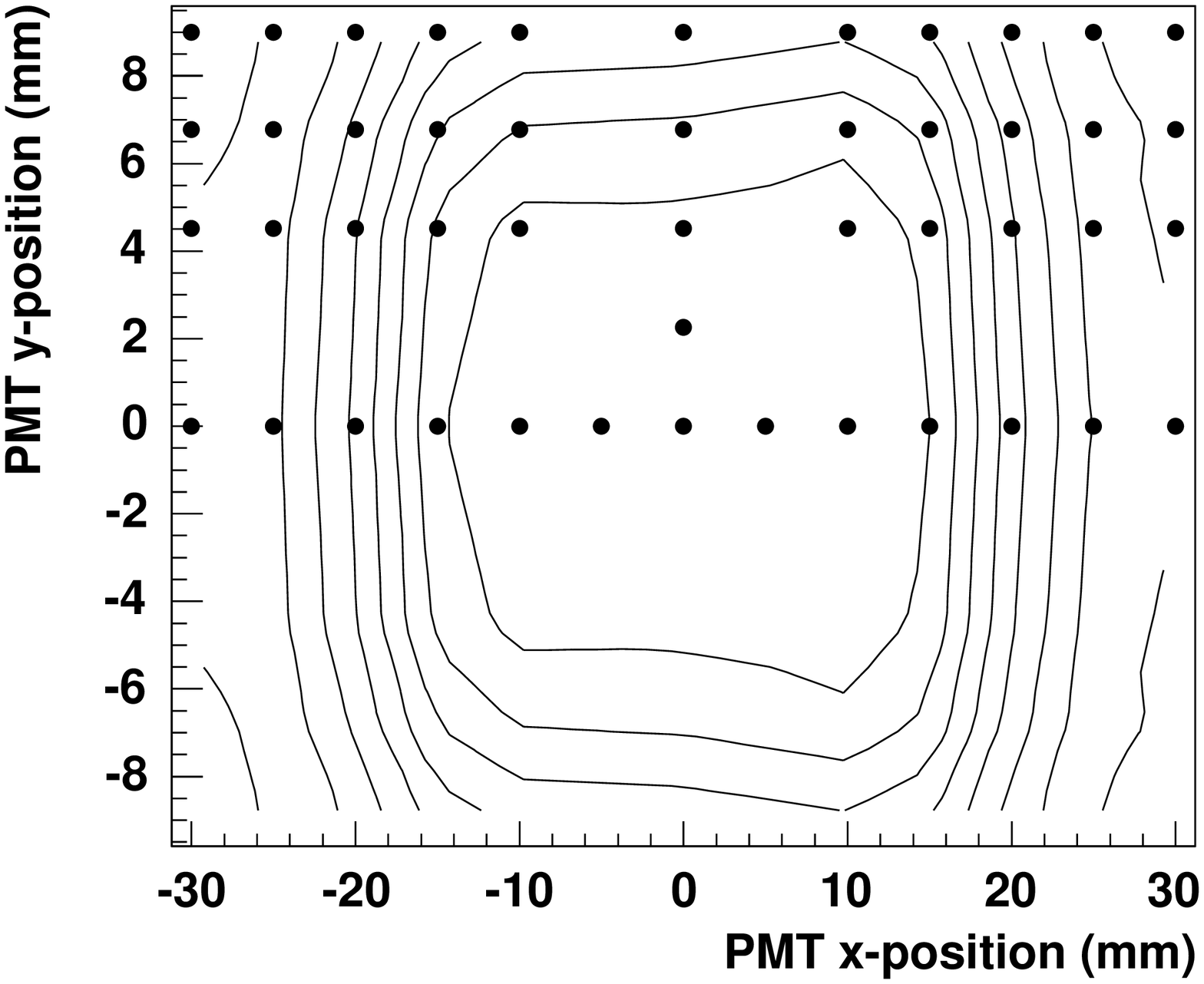}
  \caption{Measured signal time (right) and amplitude (left) as
    contour lines for the two dimensional scanning performed at a
    distance of 1\,cm from the end face of the scintillator.  The PMT
    positions for the data points are superimposed.  The steps between
    contour lines correspond to 100\,ps and $-$100\,ADC channels,
    respectively.}
 \label{fig:contour-plots}
\end{figure}

Fig.~\ref{fig:contour-plots} shows the positions of the aperture in
the detection plane as points in the x- and y-dimension. For each
point signal time and amplitude have been measured at a distance of
1\,cm from the end face of the scintillator. Only the upper half plane
was scanned due to the symmetric behaviour of all observables expected
on grounds of the geometrical symmetry. The points are superimposed on
the measured data presented as contour lines with steps of 100\,ps in
signal time (right) and steps of $-$100\,channels in amplitude (left).
The signal time variation takes place earlier in the y-direction. The
width of the scintillator in this direction is 10\,mm to be compared
to 32\,mm for the perpendicular one.

\begin{figure}
  \centerline{\includegraphics[width=0.8\textwidth]{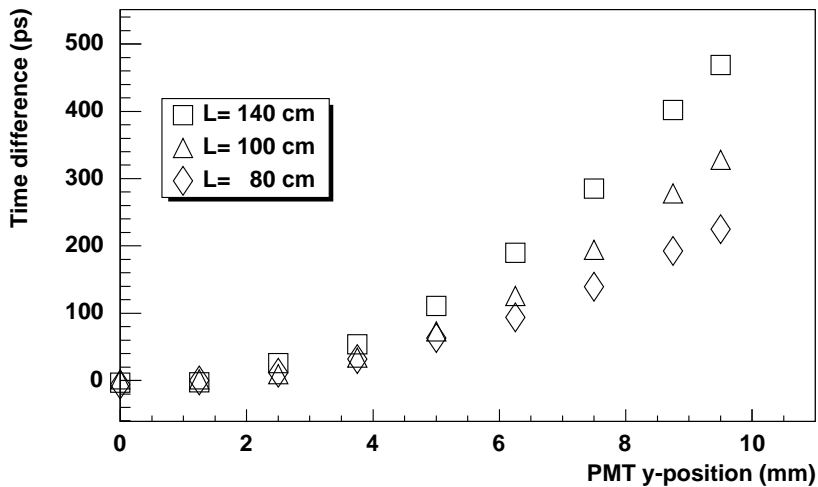}}
  \caption{Measured signal time as a function of the aperture position
    when the scintillator is excited by the UV laser at three
    different positions, $L=$ 80, 100, and 140\,cm.}
  \label{fig:timeVSpos}
\end{figure}

To exclude any amplitude dependent time shifts this experiment was
repeated for three different positions of the laser source, adjusting
the laser light intensity to equal PMT output signals for the central
aperture position. Fig.~\ref{fig:timeVSpos} shows the variation of the
signal time as a function of the position of the aperture with respect
to the centre, when the laser light is injected at $L=$ 80, 100, and
140\,cm. A signal time difference approximately linear with the
distance for a given aperture position was observed.

In a perfect scintillator the portion of the light cone that lies
inside the total internal reflection angle is transported undistorted
to the end of the slab. Since the transmission involves many
reflections, the exit angle from a real scintillator is dependent on
the sharpness of its edges and the parallelism of its faces. The
measurements showed an unambiguous correlation between signal time and
tube position and demonstrated an sufficient conservation of axial
angles. The observed correlation is not masked by a change of apparent
anode sensitivity with angle of incidence at the photocathode.
Although, the photomultiplier tube sensitivity depends on the incident
angle since the path length of a photon through the photocathode
increases like $1/\cos \theta$ giving a corresponding higher
probability for the emission of a photoelectron~\cite{Hamamatsu1999}.
The subsequent course of this electron is, however, not simple and the
cosine law can only be considered as a first approximation.  It would
imply that at non-central aperture positions amplitude losses get
partly compensated and effects of the propagation time dispersion get
emphasised. The angular variation of quantum efficiency and the
reflectance of a PMT entrance window were measured
recently~\cite{Shibamura2006}, showing a larger
photon-to-photoelectron conversion efficiency for incident angles
$\theta > 45^\circ$ than that at $\theta > 0^\circ$, however, the
reflectance is also significantly larger at these angles.

\begin{figure}
  \includegraphics[width=0.48\textwidth]{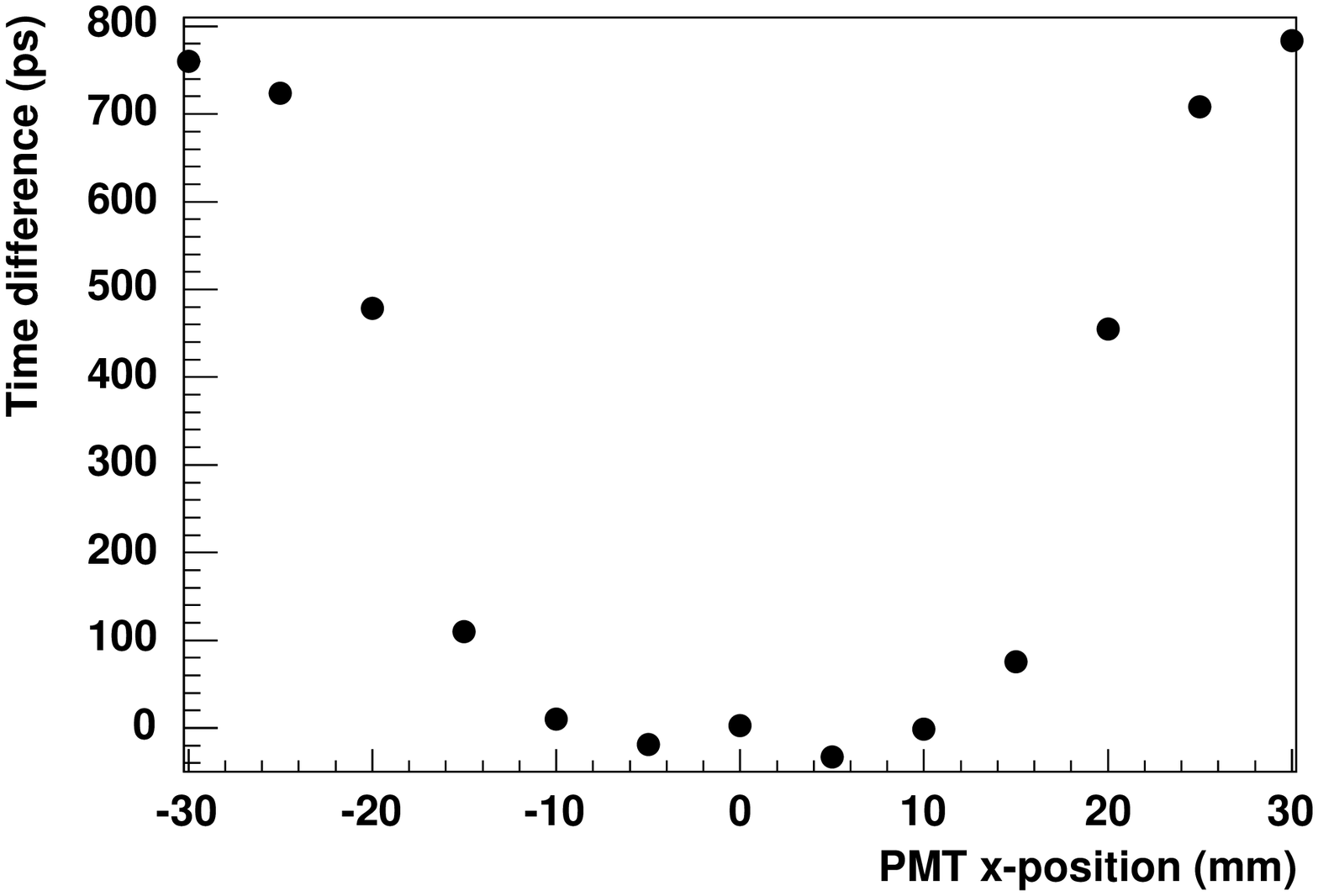}\hfill
  \includegraphics[width=0.48\textwidth]{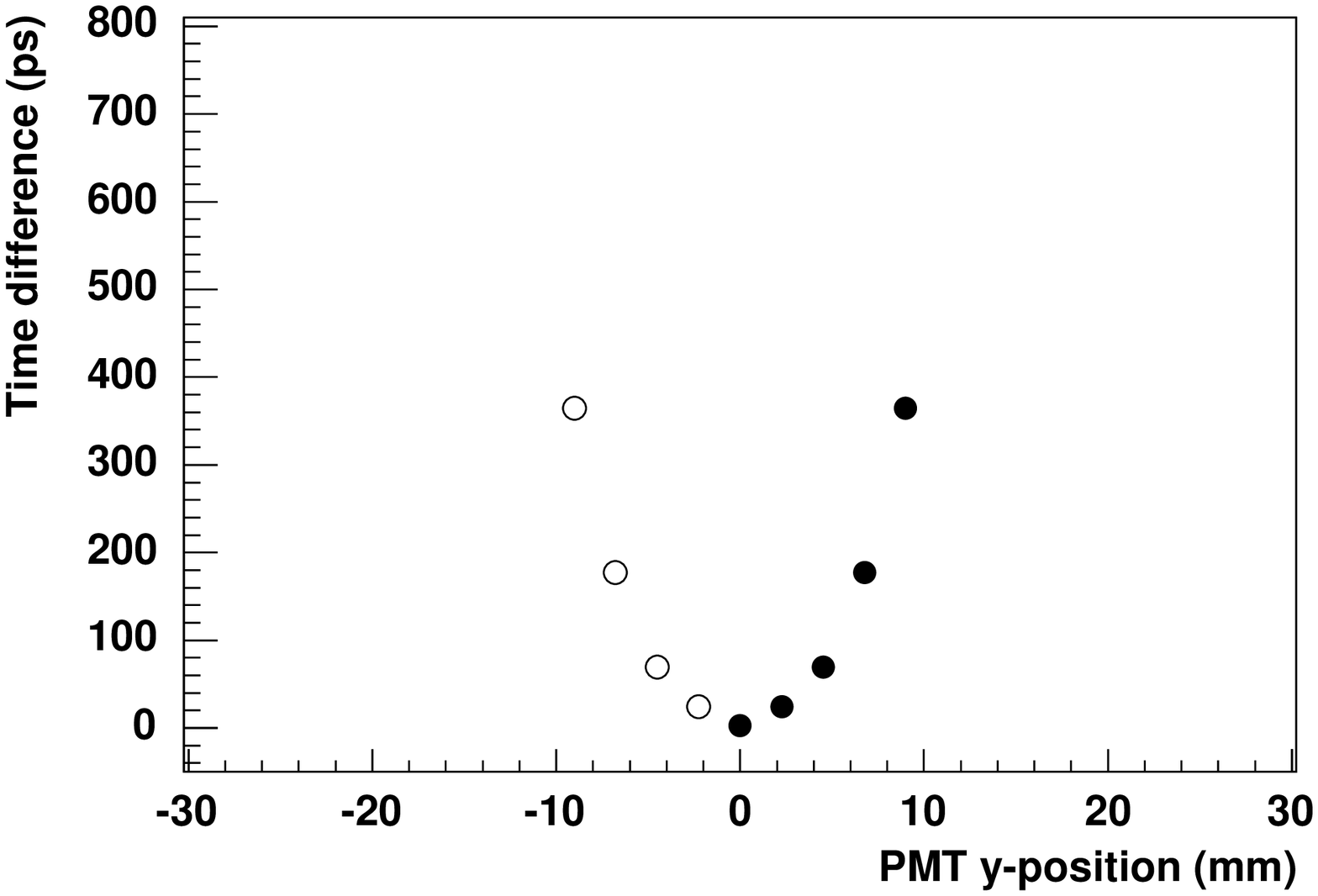}\\
  \includegraphics[width=0.48\textwidth]{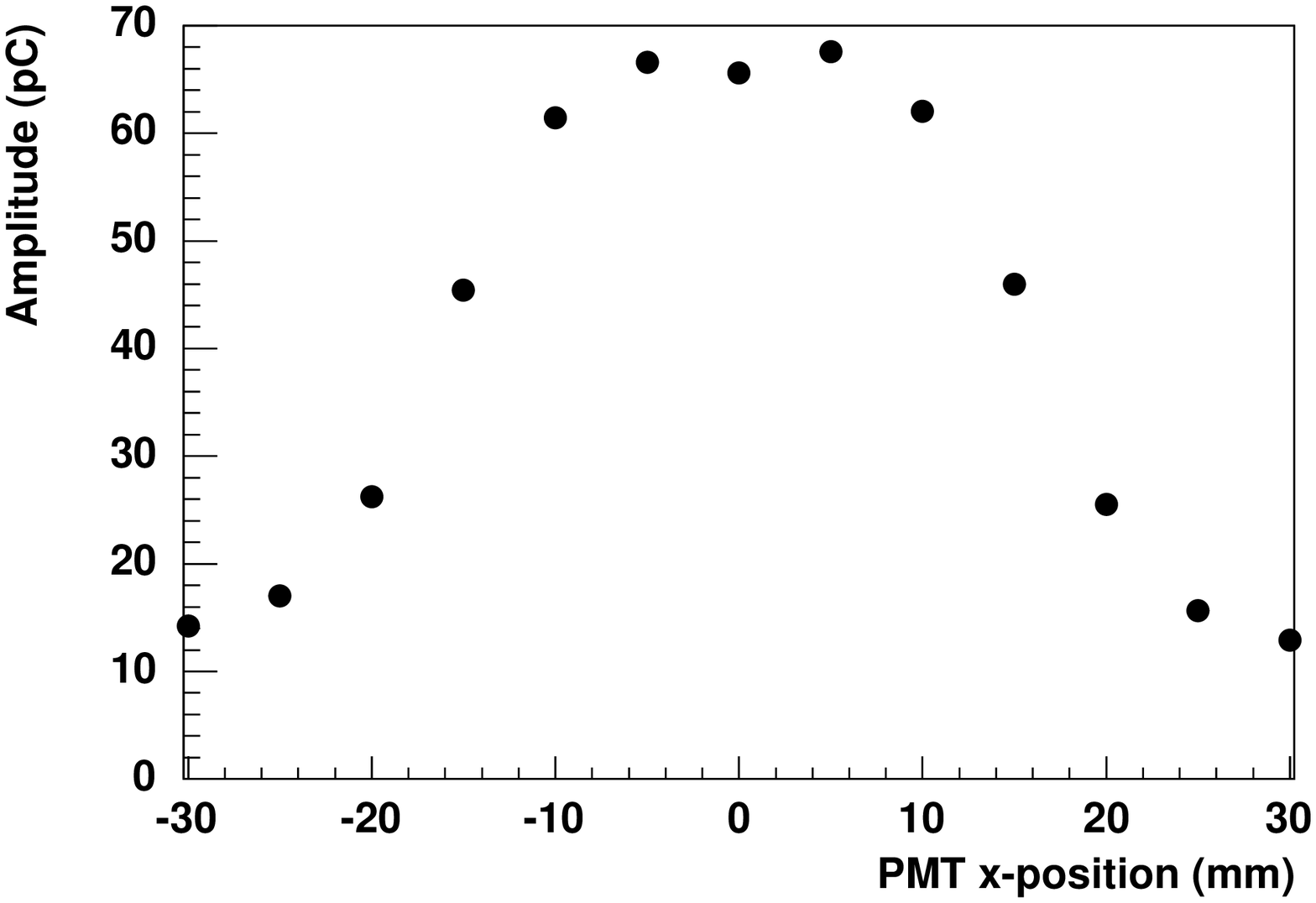}\hfill
  \includegraphics[width=0.48\textwidth]{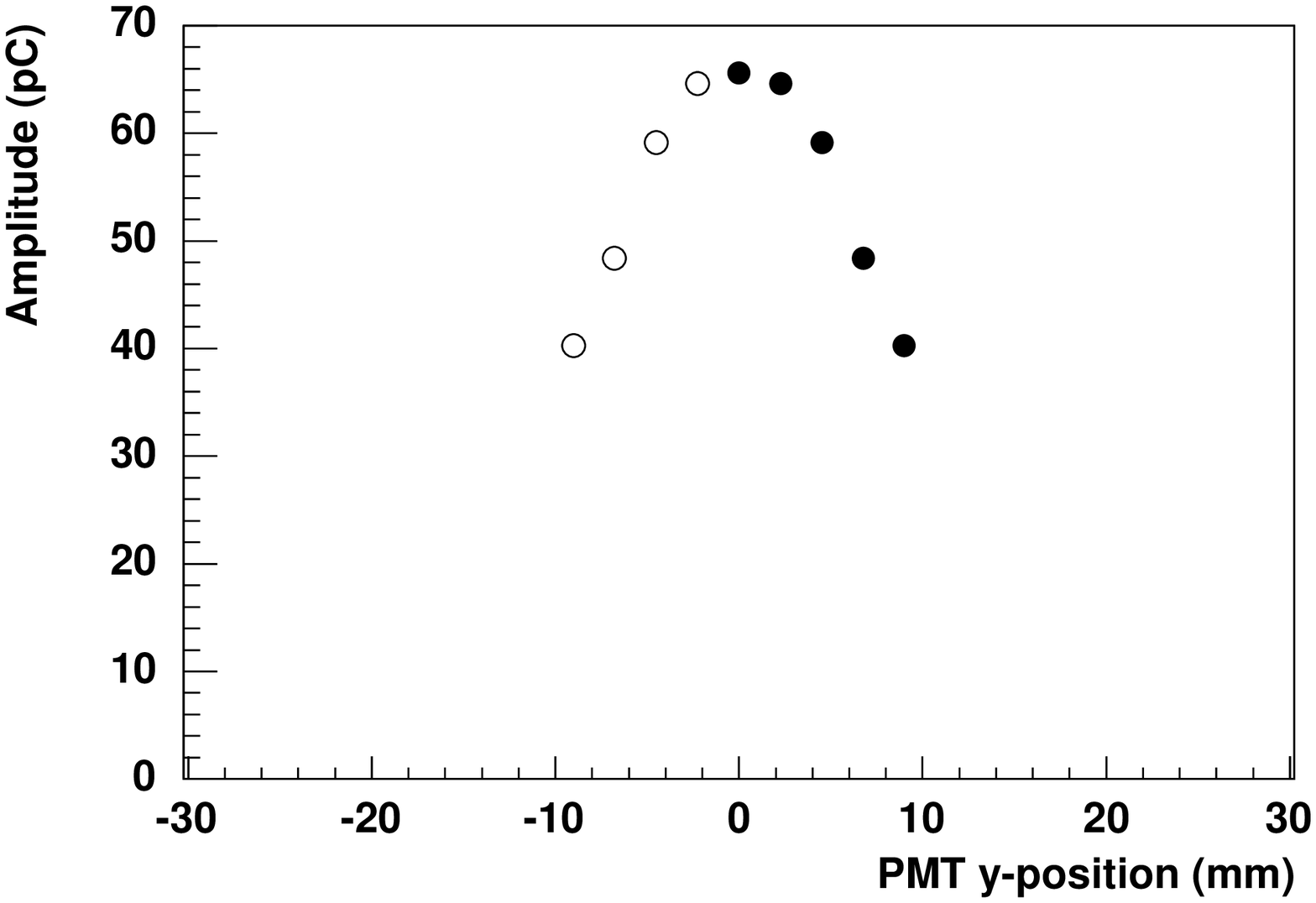}
  \caption{Measured signal time (top) and amplitude (bottom) as a
    function of horizontal and vertical aperture position. The width
    of the scintillator is 10\,mm in y-direction and 32\,mm in
    x-direction. Empty circles represent mirrored points.}
 \label{fig:measurements}
\end{figure}

\begin{figure}
  \includegraphics[width=0.48\textwidth]{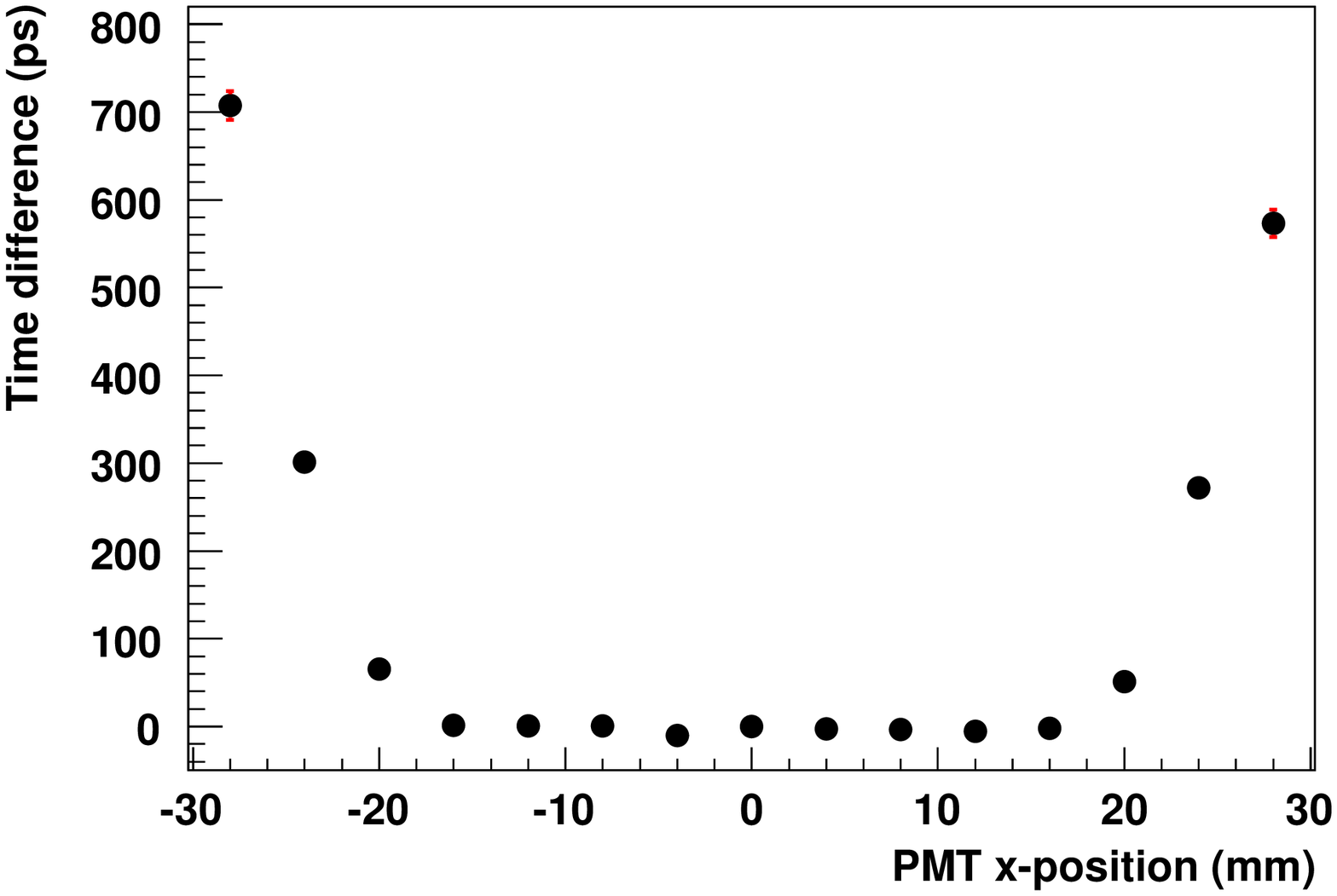}\hfill
  \includegraphics[width=0.48\textwidth]{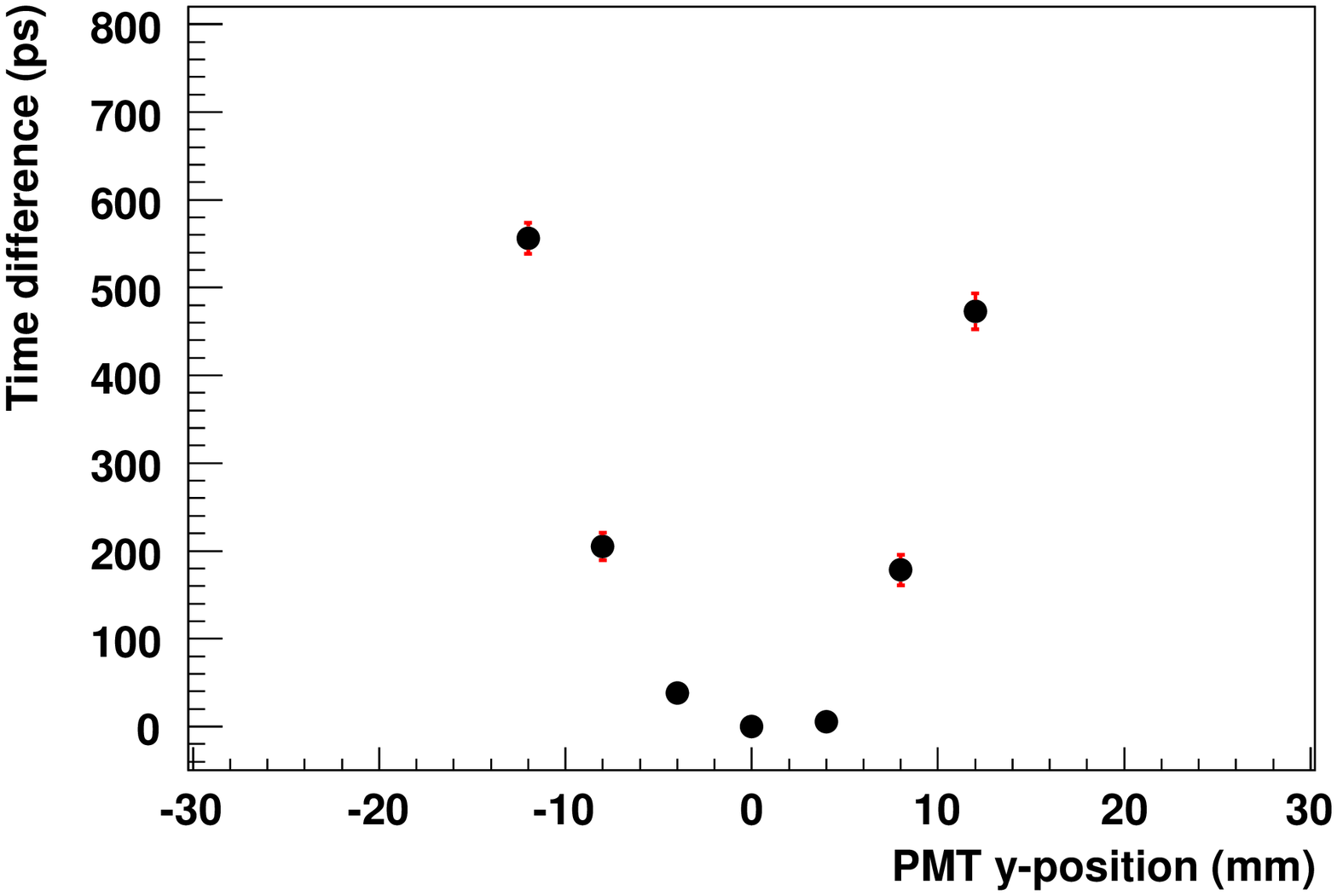}\\
  \includegraphics[width=0.48\textwidth]{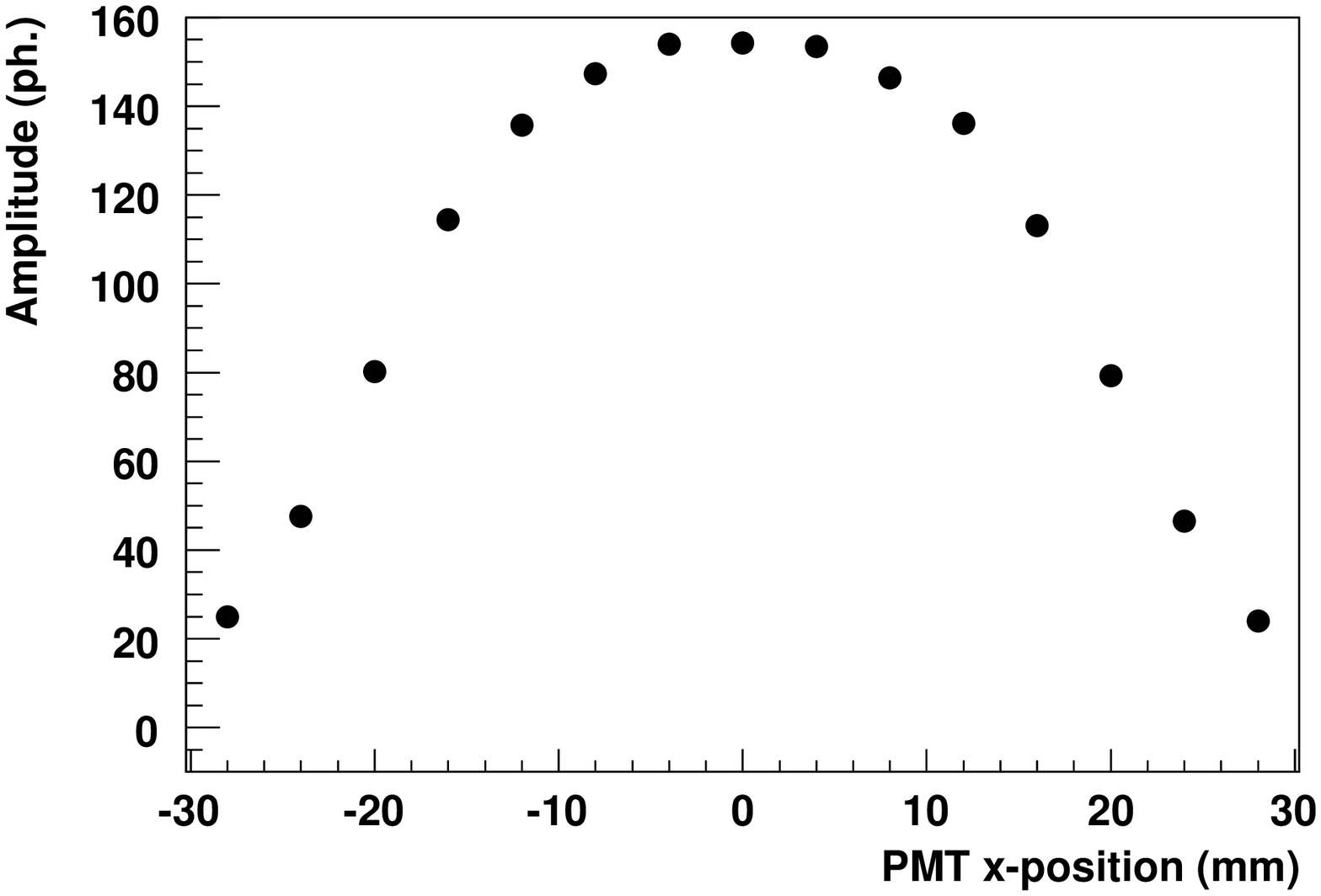}\hfill
  \includegraphics[width=0.48\textwidth]{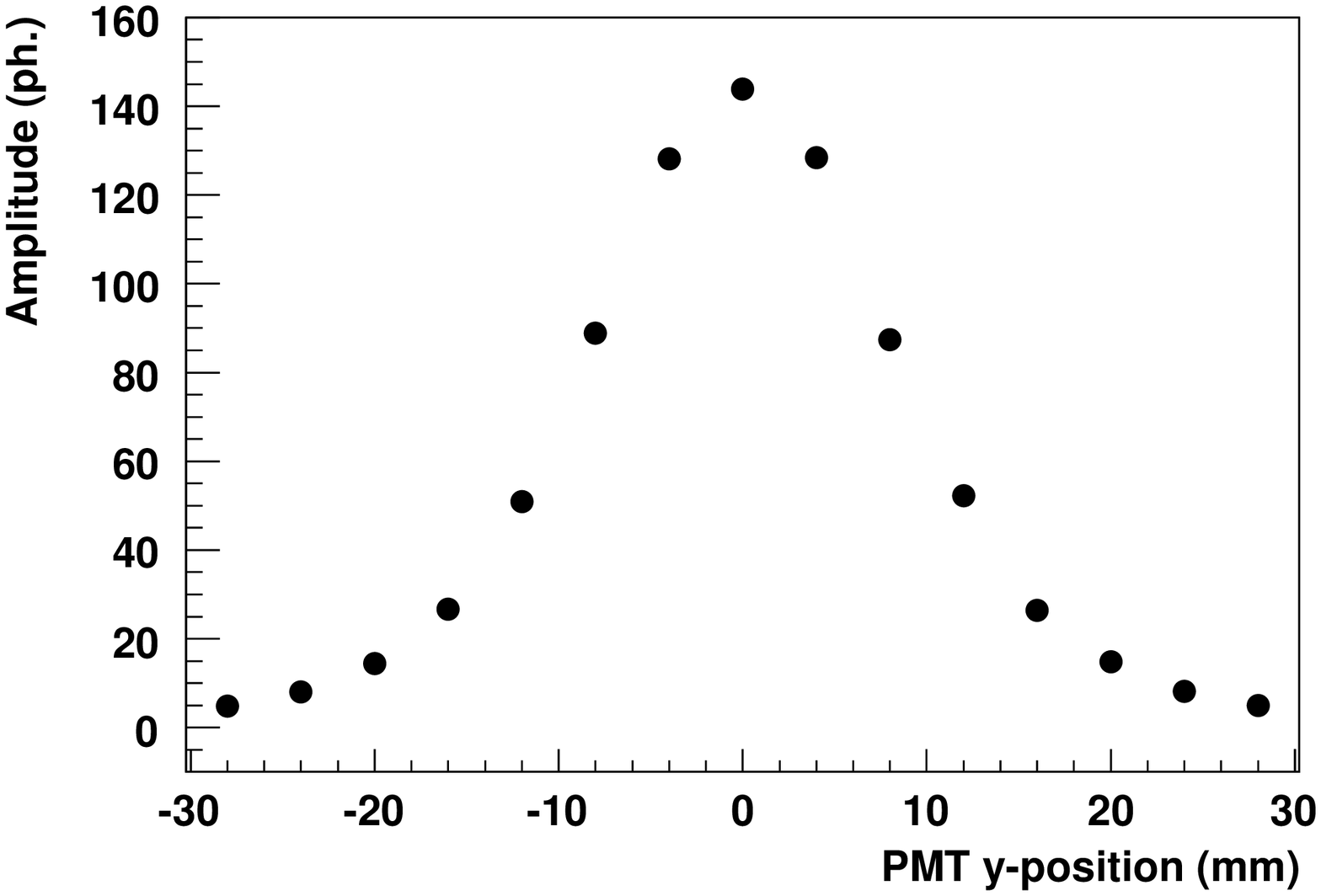}
  \caption{Simulated signal time (top) and amplitude (bottom) as a
    function of horizontal and vertical aperture position. The
    emission characteristics and the geometry of the scintillator have
    been adapted to the experiments.}
 \label{fig:simulation}
\end{figure}

\begin{figure}
  \centerline{\includegraphics[width=0.8\textwidth]{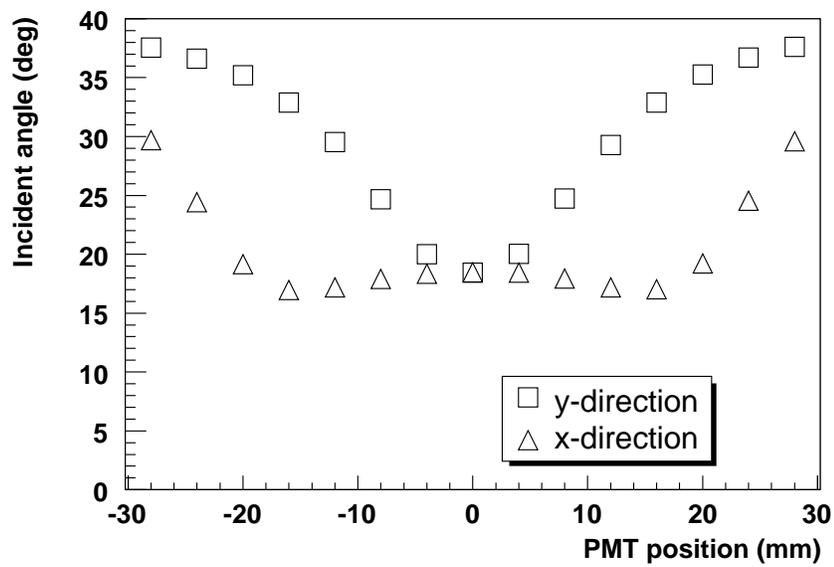}}
  \caption{Simulated average photon angle with respect to the
    detection plane as a function of the horizontal and vertical
    aperture position. The emission characteristics and the geometry
    of the scintillator have been adapted to the experiments.}
  \label{fig:angleVSpos}
\end{figure}

Fig.~\ref{fig:measurements} shows measured signal time (top) and
amplitude (bottom) as a function of horizontal and vertical aperture
position for the PMT at central position. Empty circles represent
mirrored points. An analysis of the data reveals that the observed
change in signal time is not following the change in signal amplitude.
The light propagation is simulated in detail with the Monte Carlo code
where first photoelectron timing was implemented. The observed
variation of signal time and amplitude with position was verified, as
shown in Fig.~\ref{fig:simulation}. The code was then used to simulate
the average photon angle with respect to the detection plane as a
function of the horizontal and vertical aperture position, see
Fig.~\ref{fig:angleVSpos}. The good agreement between data and Monte
Carlo together with the simulated relation between aperture position
and angle verified that the propagation time dispersion has been
observed. The measured time differences are in fair agreement with the
spread of propagation times when applying the simple $t_A - t_B= L
n/c(\cos^{-1}{\theta_A}-\cos^{-1}{\theta_B})$ formula to a pair of
simulated incident angles $\theta_{A/B}$ at two aperture positions
$x_{A/B}$ or $y_{A/B}$.

The code was also used to simulate the time resolution of a full
coverage PMT as a function of the number of photoelectrons. These
simulations served as a guide for detector
developments~\cite{Achenbach-GSISciRep05}. One example, where the
contribution of the propagation time spread $\sigma_\mathrm{prop}$
gets significant is given by the cylindrical TOF counter being
developed for \PANDA\cite{PANDA}. A thickness of only 5\,mm would
allow to mount the scintillator strips together with the DIRC
radiators, but reduces the number of photoelectrons to $\langle
N_\mathrm{pe} \rangle \simeq 100$ for particles crossing a
scintillator at $L= 1$\,m. The Monte Carlo simulation including the
finite decay time of the scintillator predicts a minimum achievable
time resolution of FWHM$_\mathrm{min}\simeq 150$\,ps, depending a
little on threshold.

These results have brought forward the idea of improving the time
resolution of a scintillation counter through position sensitive
photon detection. An analogy of this method is used successfully in
DIRC-like detectors using glass slabs for measuring the Cherenkov
angle, but was never investigated in scintillators.

Lower bounds on scintillation detector timing performance have been
developed in the past for moderate or relatively high total light
output, addressing the exponential decay of the light intensity and
non-ideal photodetectors~\cite{Clinthorne1990A}. These bounds can be
useful in determining performance sensitivity to scintillator and PMT
parameters. To the authors' knowledge the path length variation in
long scintillators as a source of additional time dispersion has not
been included in the models.

\section*{Acknowledgements}
Work supported in part by GSI as F+E project MZ/POC, by Joh.~Gutenberg
Universit\"at, Mainz, as Forschungsfonds, and by the European
Community under the ``Structuring the European Research Area''
Specific Programme as Design Study DIRACsecondary-Beams (contract
number 515873).


%

%
\clearpage


%

\end{document}